# Software Process Commonality Analysis

Alexis Ocampo, Fabio Bella, and Jürgen Münch

*Fraunhofer Institute for Experimental Software Engineering,*
*Sauerwiesen 6, 67661 Kaiserslautern, Germany*
*Tel: + 49 (0) 63 01 707 167, Fax: + 49 (0) 63 01 707 200*
*{ocampo, muench, bella}@iese.fraunhofer.de*

## Abstract

*To remain viable and thrive, software organizations must rapidly adapt to frequent, and often rather far-ranging, changes to their operational context. These changes typically concern many factors, including the nature of the organization's marketplace in general, its customers' demands, and its business needs. In today's most highly dynamic contexts, such as web services development, other changes create additional, severe challenges. Most critical are changes to the technology in which a software product is written or which the software product has to control or use to provide its functionality. These product-support technology changes are frequently relatively 'small' and incremental. They are, therefore, often handled by relatively 'small,' incremental changes to the organization's software processes. However, the frequency of these changes is high, and their impact is elevated by time-to-market and requirements change demands. The net result is an extremely challenging need to create and manage a large number of customized process variants, collectively having more commonalities than differences, and incorporating experience-based, proven 'best practices'. This paper describes a tool-based approach to coping with product-support technology changes. The approach utilizes established capabilities such as descriptive process modeling and the creation of reference models. It incorporates a new, innovative, tool-based capability to analyze commonalities and differences among processes. The paper includes an example-based evaluation of the approach in the domain of Wireless Internet Services as well as a discussion of its potentially broader application.*

## 1. Introduction

Survival in today's highly dynamic business environments requires that organizations continuously adapt their processes. Success and growth — rather than mere survival — require that this adaptation be rapid enough to realize the competitive advantage offered by new business opportunities. Business models must be rapidly changed or newly developed; the organization's work force must be quickly updated and trained. Most challenging, however, is rapid adjustment to changes in the organization's *process-support technology*. For organizations providing software products (*software organizations*), this includes the technology used to develop their products as well as the technology the products must control or use to provide their functionality. In many software-dependent areas, for example the Wireless Internet, single process-support technology changes are small and incremental, but quite frequent. The result of such changes can be a set of processes that vary in relatively minor ways. However, organizations suffer severe problems when such changes are introduced arbitrarily, irrationally, and in uncontrolled ways. They could, for example, be the cause of drastic deviations from project plans.

One way to control the proliferation of variations and its attendant risks is to carry forward knowledge about what worked and what did not work in the past. In other words, software organizations working in this area must be agile, but this agility should be based on prior experience rather than being merely based on intuition (Boehm 2002). Process changes should be to the best possible extent beneficial for the organization's future work.

This raises the question: How can a software organization cope with product-support technology changes by rapidly creating customized software development processes containing proven 'best practices'?

The rationale for our work is that understanding an organization's current and past practices, describing the processes underlying these practices, and being able to identify variations and reasons for variations will certainly help software organizations address this question.

The basis for our approach is the creation of customizable, domain-specific process models (i.e., reference process models) through the bottom-up identification of process variations. The overall approach can be seen in Figure 1, and is described with details in (Becker-Kornstaedt *et al*. 2002). The mentioned article describes a validated method on how to gain process knowledge for an upcoming field fast and incrementally. It is not the scope of this paper to detail the overall approach. Briefly, the method can be used for designing an adaptable software development process based on existing practices from related domains, industrial piloting, and expert knowledge. Its main steps are:

- *Set up pilots* - Suitable pilot projects are determined and organized.
- *Perform pilots* - The pilot projects are conducted.
- *Observe and model processes* - The processes as performed in the pilot projects are observed and modeled.
- *Identify and evaluate processes and practices from related fields* - This information will be used to complete the reference process model where it is incomplete.
- *Analyze commonalities and differences* - Commonalities and differences between the different process models are analyzed in order to identify process variants and justifications for them. This must recognize differences in the application domain as well as goals and contexts of the pilot projects.
- *Create comprehensive process model* - The models for the processes used in the pilot projects as well as practices and processes from related fields are integrated to create a comprehensive process model (Ocampo *et al*. 2003). The resulting comprehensive process model can be seen as a *reference process model*, because it is intended to be used as a reference for developers and managers that provides a starting point for developing a customized process meeting the requirements for a set of product-support technologies.

The focus of this paper is to present a tool-supported technique for performing the activity *analyze commonalities and differences* (part of the overall approach), which can be helpful in practical situations where software organizations must compare a set of process models in a systematic way, in order to understand their context-dependant variations.

Capabilities for identifying best practices and process variations might be valuable for software organizations in other areas, too, such as:

- Process measurement: Metrics that reflect process similarities and differences could be important for guiding process improvement.
- Process training: Identifying the gaps between desired and actual processes could improve workforce

training.
- Tailoring guidance: Notations for describing common and alternative process parts could support tailoring.
- Outsourcing: Commonality analysis could provide a basis for integrating processes between an outsourcing organization and organizations it outsources to.
- Executive decision making: Notations for assigning value to variations can be important for managers of software organizations for deciding how to change a software project so that it may proceed more effectively and efficiently.

The following sections discuss the background for our work, describe the details of the technique we have developed, provide a preliminary validation (in terms of an example of its use), and discuss possible future work.

## 2. Background

The following section presents commonality analyses performed in related fields as well as a description of the context of the work.

### 2.1. Related Work

In the database world, the problem of integrating schemas of existing databases from the perspectives of different users (database schema integration) is addressed by (Batini *et al*. 1992). Products from this database integration are: a global database schema, data mapping from global to local databases, and mapping of querying transactions from local to global databases. Semantic relationships between database schema X1 and database schema X2 are defined as: identical, equivalent, compatible, and incompatible. The schemas are analyzed and compared in order to uncover conflicts. Any situation where the representations of X1 and X2 are not identical is considered to be a conflict between X1 and X2. The representations of the schemas are used to compare them, but there is no defined method to do this comparison.

Integration of design specifications has been examined by (Feather 1989), (Leite and Freeman 1991), (Robinson 1989). These approaches have in common that they integrate pairs of specifications and use specification formalisms, and that their goal is to reduce the complexity of the global specification. The analyst compares components of both specifications and declares them equivalent or not. A special formalism is used in order to conclude when a component X1 is equivalent to component X2. Conflicts are uncovered when ambiguities and inconsistencies are detected between pairs of specifications. Negotiations are needed between developers in order to identify and resolve conflicts. Once the integration has been accomplished, there is no way to extract the original views from the final specification, which is not the case with the technique presented in this article.

In the product line world, identifying commonalities and differences is an accepted, wide-spread practice when comparing systems (Coplien *et al*. 1998). Usually, common elements are reused and variations are hidden, in the most appropriate way, in order to produce a family of products. In order to understand the extent of commonality and variability in a family of products, the proposed steps are:

- *Establish the scope* - The collection of objects under consideration.
- *Identify commonalities and variations* - Similar attribute values across the family members are identified. Variants of the attribute values are identified. The attribute values justify the variants.
- *Bound variations* - A range of values for the variants is defined.
- *Exploit commonalities and accommodate variations* - The results of the commonality analysis are grouped into procedures, inheritance, and parametric polymorphism.

In the process modeling world, there exist some approaches to integrating partial process models (*views*) into a descriptive process model when persons covering different roles describe their perspectives of a large software process within one single organization (Turgeon and Madhavhji 1996), (Verlage 1998). In these approaches, variations are often seen as inconsistencies or as imprecision, and therefore, trigger questions that lead to a review of the views that will eliminate these inconsistencies. The final goal is to obtain a multi-view-consistent comprehensive process model. In our approach, some of the rules discussed in (Verlage 1998) are adapted to the SPEARMINT® environment (Becker-Kornstaedt *et al*. 2000) and applied in order to create a model with common best practices and variations to be used as reference within a specific domain.

In the process modeling world, creating reference models is often done in a top-down fashion using prescriptive process models. Prescriptive process models describe how a product *should* be developed. Prescriptive process models are generic (i.e., do not define specific approaches to carrying out activities), and do not describe a company's actual processes.

The commonality analysis technique we propose can be seen as analogous to the commonality analysis of products in product line approaches. It relies on descriptive, rather than prescriptive, process modeling to create a reference model, utilizing several capabilities found in multi-view modeling approaches (such as rule-based comparisons).

## 2.2. Context

The technique was developed and evaluated as part of the WISE (Wireless Internet Services Engineering) project. The project aimed at producing integrated methods and COTS components (commercial off-the-shelf software) and open source to engineer services on the Wireless Internet. The components include a service management component and an agent-based negotiation component. Two pilot services, i.e., a financial information service and a multi-player game, were developed by different organizations. The project lasted 30 months and an iterative, incremental development style was applied: three iterations were performed of roughly 9 months each. In iteration 1, a first version of the planned pilot services was built using GPRS (General Packet Radio Service). At the same time, a first version of methods and tools was developed. In iteration 2, a richer second version of the pilots was developed on GPRS, using the first version of methods and tools. In parallel, an improved second version of methods and tools was developed. In iteration 3, the final version of the pilots was developed on UMTS (Universal Mobile Telecommunications System), using methods and tools from the second iteration. Also, a final version of methods and tools was developed. One of WISE's objectives was to develop a reference process model that may be used by software organizations for creating Wireless Internet Services. In order to achieve this objective, it was decided to use the empirical approach, i.e., the observation of realistic pilot projects, techniques, and processes, described previously in the introduction section of this article (see Figure 1). The SPEARMINT® tool has been used in order to document and manage the pilots' descriptive process models. The decision for using it relied on previous industrial case studies with the tool, where its value has been proven (Becker-Kornstaedt *et al.* 2001). The following section will present in detail the technique for identifying commonalities and differences between processes.

## 3. Technique

The technique proposed here is based on the assumptions that the same notation must be used to describe the process models to be compared, and that the level of abstraction of the process models to be compared must be similar.

In order to perform a commonality analysis, the models must be rigorous. This may be achieved, for example, by using electronic process guide (EPG) capabilities with graphical views (Becker-Kornstaedt *et al*. 2000).

In order to validate the proposed commonality analysis technique, this technique was performed both manually and by using a specifically developed tool, SPEARSIM.

### 3.1. Manual Commonality Analysis

By using electronic process guides (EPGs), the process engineer can identify whether two parts of different process models are similar. For example, looking at Figure 2, it can be seen that there are two processes with the same name (Test acceptance), both in pilot 1 and in pilot 2. Additionally, the structure of the processes is similar. In both cases, there is a similar product to be tested as input (Release system and Tested product), and a similar report as output (Defect list and Test report). Once similar process parts are identified, the process engineer reads the definition of the processes and products related to these parts. After reading and analyzing the descriptions, the process engineer makes the assumption that two or more processes or sub-processes are similar or different.

The process engineer has to check the descriptions of processes, products, roles, and tools in order to establish an assumption that they are similar. The next step is to check the assumption by reviewing the identified commonalities with the process performers, that is, the observed developers, in order to obtain a common agreement on the commonalities, i.e., establishing facts. If the activities are not similar, then the next step is to find possible reasons for the variation. The reasons can usually be found in the context of the process, which is described in the characterization vector (see Table 1). The characterization vector describes the environment in which the process model was elicited. The characterization vector shown in Table 1 is the result of the activity *set up pilots* (see Figure 1).

### 3.2. Tool-supported Commonality Analysis

The tool SPEARSIM has been designed to support a process engineer in comparing large and complex processes. SPEARSIM is implemented as a plug-in for SPEAMINT® (Becker-Kornstaedt *et al*. 2000). The tool analyzes the similarity of two process models using a set of rules, which are derived from the heuristics applied by Verlage in the context of the Multi View Modeling Language (MVP-L) (Verlage 1998). The rules formalize different similarity aspects that may occur between entities of two process models and, consequently differ, in their degree of complexity: on the one hand, simple rules can be used to compare entity names (such as process or product identifiers) and help to identify synonyms and homonyms; on the other hand, more complex rules can be used to compare the aggregation structure of products and processes. Figure 3 shows an overview of the rules defined and their dependencies. The dependencies, represented by arrows, show that the computation of complex similarities rests upon data computed by simpler rules.

In the following, the individual rules are discussed in more detail:

**Name** – This rule is applied to compute the similarity of products/processes based on the similarity of their names. This rule computes text similarity according to the Levenshtein distance (Levenshtein 1966). The Levenshtein distance (LD) is a measure of the similarity between two strings, which we will refer to as the source string (s) and the target string (t). The distance is the number of deletions, insertions, or substitutions required to transform s into t. For example, If s is "test" and t is "test", then $LD(s, t) = 0$, because no transformations are needed. The strings are already identical. If s is "test" and t is "tent", then $LD(s,t) = 1$, because one substitution (change "s" to "n") is sufficient to transform s into t. The greater the Levenshtein distance, the more different the strings are. This rule provides the basis for the entire computation at the beginning of the analysis process.

**SC** (Structure Compatibility) - The SC rule can be applied on two sets of processes or products. The value computed by SC represents the degree of homogeneity of the two sets, i.e., how well the entities of one set match the entities of the other set. For example, given two sets $A = \{a, b, c\}$ and $B = \{d, e, f\}$ where $b$ and $f$ are the only identical entities between the two sets, i.e., the number of matches is $m = 1$ and the maximal number of matches is $n = 3$, the similarity value returned by SC is computed as: $\frac{m}{n} = \frac{1}{3}$.

**PcH** (Process Hierarchy) - The PcH rule computes similarities between processes by analyzing the hierarchy of their sub-processes. Since a comparison of the entire aggregation tree can become very complex, the computation is only concerned with the first three hierarchy levels of the tree structure. The PcH rule extracts the greatest Name similarity among the sub-processes of two processes, for example, $PcH(p_1, p_2)=1$, if *{"write test cases", "implement test cases", "run test cases"}* are the sub processes of $p_1$ and *{"code test cases", "run test cases"}* are the sub-processes of $p_2$, since *Name("run test cases", "run test cases")*=1.

**PdS** (Product Structure) – The similarities between two processes are computed by the PdS rule resting upon the homogeneity of the sets of products the two processes access, i.e., the products they produce, consume, or modify. The PdS rule applies the SC rule: continuing with the example discussed under the SC rule, *PdS($p_1$, $p_2$)*=$\frac{1}{3}$=*SC(A,B)* holds if *A* and *B* are the sets of products accessed by the processes $p_1$ and $p_2$, respectively.

**PcS** (Process Structure) - The PcS computes similarity assumptions between two processes resting upon the homogeneity of the sets of sub-processes they aggregate. The PcS rule, like the PdS, applies the SC rule. In this case, *PcS($p_1$, $p_2$)*=$\frac{1}{3}$=*SC(A,B)* holds if *A* and *B* are the sets of sub-processes of the processes $p_1$ and $p_2$, respectively.

**PcM** (Process Model) - The similarity values are computed by the PcM rule by building a weighted sum of the rules PdS, PcS, and PcH,

i.e., $PcM(p_1, p_2) = w_{PdS} \cdot PdS(p_1, p_2) + w_{PcS} \cdot PcS(p_1, p_2) + w_{PcH} \cdot PcH(p_1, p_2)$ where

$w_{PdS} + w_{PcS} + w_{PcH} = 1.0$ are the weights set by the process engineer to influence the relevance of the related rules in the computation.

In order to influence certain aspects of the models assumed to be relevant according to the given context, the process engineer is able to influence the importance of the different rules by setting parameters (weights) in the tool.

Once all the weights are set, the process engineer can trigger a first computation of similarities to be performed by the tool. The tool quantifies the similarity of two process models using the rules and shows them to the process engineer in the form of similarity assumptions.

The process engineer may need to read the descriptions of the compared parts of the process, in order to have an adequate basis upon which to understand these assumptions. If this is not enough, the process engineer should interview the process model owners, (e.g., in the WISE project, these were the developers) in order to better understand the assumptions made by the tool.

The process engineer converts assumptions into facts by accepting or rejecting the assumptions computed by the tool. A fact is represented by either an equal symbol (=) or a difference symbol (≠). Once the facts are established, the tool can use them to re-analyze the two models and present a new set of assumptions to the process engineer, who decides whether to continue with a new iteration by establishing new facts or whether to stop the comparison. In order to achieve a sharper picture, the process engineer can trim the weights once again, and try to get most of the greatest similarities computed for the pairs expected to be identical, most of the lowest similarities for the pairs expected to be completely different and, at the same time, maximize the difference between great and low similarities.

Figure 4 shows an excerpt of the table of commonalities: In this case, the process engineer has turned all the assumptions into facts. Activities like *elicit first requirements* and *gather requirements* were concluded to be similar. On the other hand, activities like *documenting* from pilot 1 were different from any activity of pilot 2. The resulting table of commonalities can be used in reviews with developers to build the comprehensive process model. Expected great similarity values indicate evidence of a common path between the compared processes. Unexpected similarity values characterize the most interesting pairs of process entities, since they could indicate variations between the compared processes.

## 4. Validation

This section presents a preliminary validation (proof of concept) of the proposed approach by providing an example of its use. The following example was performed in order to validate the suitability of the semi-automatic commonality analysis in a real environment (like the WISE project), by comparing its results with the results from the manual commonality analysis.

Within the WISE project, two different software development lifecycles applied by two different organizations, with 10 and 12 sub-processes, respectively, were compared. In a first step, a manual comparison was performed and pairs of similar process parts were documented as shown in the examples appearing in Figure 2. In a second step, the SPEARSIM tool was used. Similarity facts between products were established by the process engineer according to the content and purpose of the documents manipulated by the different processes.

In a third step, a computation was performed in order to analyze commonalities between the processes within each phase. Finally, another computation was performed in order to analyze commonalities between the different phases of the two development processes.

Figure 5 presents a view of the similarity values generated by SPEARSIM, showing the similarities between phases.

The phases of both models were settled in a chronological order in the diagram. As a consequence, the greatest similarities were expected along the main diagonal (highlighted by the ellipse). Parts of the diagram not matching the expectations are an indicator of either variations in the two processes (in the case of low similarity values among the main diagonal) or too optimistic tool computations (in the case of great similarity values in other areas of the diagram). Figure 5 shows the greatest commonalities in the requirements as well as in the test phases of the two development processes. These results were also observed in the manual analysis, an example of which can be seen in Figure 2, where basic activities of the testing phase of both pilots' processes were declared similar. A mismatch of the development phase (pilot 1) and the coding phase (pilot 2) shows where to expect the greatest differences between the two development processes. The main reasons for the differences were found in the maturity of the software development organizations responsible for the development of the pilot services as well as in the different final products, a WML (Wireless Markup Language)-based information system in the case of pilot project 1, and a distributed game implemented in Java in the case of pilot project 2. The great similarities between the requirements and the test phases of the two processes, respectively, indicate an example of an optimistic similarity computation due to the underlying similarity of the products manipulated by these phases (i.e., products concerning requirements).

Figure 6 shows the similarities computed between the underlying processes, which are arranged on the axes in a chronological order. The weights were chosen in order to consider only the structure of the products accessed by the constituent processes: As the processes were almost not aggregated or the aggregations were not comparable further, an analysis of their structures was avoided. Although a more complex situation is given here, in this case, most of the greatest similarity values are also arranged, as expected, along the main diagonal of the diagram.

The life cycle model applied for Pilot 1 does not include any process for the planning of tests. The unexpected great similarity values between the process *Pilot 2 - plan tests* and the processes *approve requirements, design technical infrastructure, specify requirements, develop pilots, design web site, and create technical infrastructure* in Pilot 1 can be explained by the similarity of the requirements-related documents accessed by these processes.

The whole process lasted approximately 30 hours from the identification of similar and different parts in the EPGs until the identification of reasons for variations. The whole commonality analysis performed with the help of SPEARSIM lasted about 15 hours. One fact observed in this proof of concept was that the time spent on the manual analysis was about one half of the time spent with the help of the semiautomatic tool. However, we cannot rely on this for drawing any conclusion, especially because performing first the manual analysis and then the automatic analysis has an impact that remains uncertain. This is the subject of further research. Even though more data should be collected in future research work, it was observed that the tool, through its visualization capabilities, offers a comprehensive map of commonalities and differences that certainly helps to visualize where processes are more different. This has been pointed out by the process engineers who performed the example as an important factor that makes the work easier when using the tool support.

The previous discussion suggests that the similarity values delivered by the tool are true. However, further research should concentrate on developing metrics for measuring how accurate these similarity calculations are.

## 5. The Reference Process Model

Although the similarity values provided in the previous section may appear obvious, the objective of this commonality analysis went further, because we intended to uncover process similarities across different pilot projects (i.e., final products) in one domain (e.g., Wireless Internet Services), which we can then nominate as 'best practices' that should appear in any customized version of the reference process, as well as variations to be taken into account under special context characteristics. Having evidence of common practices across projects certainly provides a basis for making the assumption that such a practice can be declared a 'best practice'. However, this is not always the case. Variations can as well be nominated as 'best practices' by developers. In the end, developers are the responsible of accepting or rejecting the nomination of a 'best practice', and introducing it into the reference process model. The following is an example of how the results from the commonality analysis were used for creating the reference process model in the context of the WISE project.

Table 2 presents two symbols needed to understand the reference process model descriptions. The new symbols are used for grouping those sets of processes that are considered optional or alternative in the reference process model. Those sets of processes that were considered similar after the commonality analysis were named basic/common activities, and were not grouped into these boxes. Processes that were grouped into the optional boxes were the following:

- Processes for which no similar process was found in the process model used for comparison.
- Processes that were not followed by developers during the previous development iteration. Usually, these were processes that had to be skipped due to time constraints, but that were considered important by developers.

Processes that were grouped into the alternative boxes were the following:

- Processes with similar purposes that were found in the process model used for comparison but whose steps, tasks, or means to fulfill that purpose were different.

Figures 7 and 8 illustrate the use of the optional and alternative boxes in a product flow graph.

Figure 7 is one snapshot taken from the main view of the WISE reference process model. This snapshot shows merged experience-based, proven common practices and variations. Based on the results of the commonality analysis and with constant feedback from pilot partners, the process engineer merged, for example, the requirements phases from both processes, into one common part in the reference process named *requirements phase*. The same was done with the *coding phase* and *testing phase*. On the other hand, pilot 1 did not present evidence of performing the activities *build test framework* or *plan tests,* therefore, it was considered different by the process engineer and pilot partners and merged into the reference process model as a variation of the process.

Figure 8 presents two alternative boxes corresponding to the integration/releasing activities of both pilots. By simply looking at the activities on the EPGs, they could be declared common, because of their similar names and structures. However, after reading the compared activities descriptions, they were kept separated and included in the reference process model as alternatives. For example, the activity *integration testing* present in the ALT1a box was performed between developers and market experts and therefore, was quite different from the *integrate code* activity from ALT1b, where only developers dealt with the integration. It can also be seen that the documenting activity was inserted as optional inside ALT1a. This is due to the fact that documenting was part of the process model of one of the pilots, whereas nothing similar was found in the process model used for comparison.

## 6. Summary and Outlook

The technique presented in this article is helpful for managing the comparison of large, complex processes, and the rules are applicable for processes in the same organization as well as in different organizations. Nevertheless, in the course of the exercise discussed in the previous section, we noticed that some assumptions made by the tool were not as concise as we expected. This suggests that it would be good to perform further research on similarity computation rules, in particular, rules followed intuitively by process engineers, in real practice, to determine best practices and process variations. Also, some additional process attributes (e.g., measures, estimates, standards applicable, pre-/post-conditions, etc.) are not considered in the actual version of the tool. Future research may then address the following questions: What are other rules to determine process similarity and dissimilarity? When should these rules be applied, and when not? Which degree of process complexity requires automated similarity analysis? Which metrics can be applied for measuring accuracy of similarity values? Which other values for establishing facts (e.g., very similar, similar, low similarity, no commonality) could be used?

Regarding the creation of the reference process model, it was revealed that appropriate notations for describing generic process knowledge (i.e., adaptable process models and adaptation rules) are needed. Optional product flows, for instance, are difficult to represent in existing notations. An attempt to represent this information was developed and used for describing the Reference Process Model. Experiences on understanding this approach are still being collected.

The resulting reference process model together with the characterization vector describing the context in which the model was created can be used by software development managers or software process managers for understanding, analyzing, defining a strategy, or defining a process in order to develop Wireless Internet

Services. In fact, in the context of the WISE project, this is exactly what was done. After each of the three iterations where pilot services were developed, the processes were compared through a commonality analysis, and the reference process model was updated (or created, in the case of the first iteration). The resulting reference process model was then used as input for the next iteration.

## Acknowledgments

This work has been funded by the European Commission in the context of the WISE project (No. IST-2000-30028). We would like to thank the WISE consortium, especially the pilot partners, for many fruitful and interesting interactions. We would also like to thank Dr. William E. Riddle for his valuable comments on the article, and Sonnhild Namingha from the Fraunhofer Institute for Experimental Software Engineering (IESE) for reviewing the first version of the article.

## References

Batini C, Ceri S, Navathe S.B. 1992. Conceptual Data Base Design: An Entity-Relationship Approach. Benjamin Cummings: Redwood City.

Becker-Kornstaedt U, Boggio D, Münch J, Ocampo A, Palladino G. 2002. Empirically Driven Design of Software Development Processes for Wireless Internet Services. PROFES 2002: Proceedings of the Fourth International Conference on Product-Focused Software Processes Improvement. Lecture Notes in Computer Science: 351-366.

Becker-Kornstaedt, Ulrike; Neu, Holger; Hirche, Gunter. 2001. Software Process Technology Transfer. Using a Formal Process Notation to Capture a Software Process in Industry. EWSPT'2001: Proceedings of the 8th European Workshop on Software Process Technology: 63-76.

Becker-Kornstaedt U, Scott L, Zettel J. 2000. Process Engineering with Spearmint/EPG. ICSE'00: Proceedings of the 22nd International Conference on Software Engineering: 791.

Boehm B. 2002. Get Ready for Agile Methods, with care. IEEE Computer 35(1): 64-69.

Coplien J, Hoffman D, Weiss D. 1998. Commonality and Variability in Software Engineering. IEEE Software 15(6): 37-45.

Feather M.S. 1989. Detecting Interference when Merging Specification Evolutions. Proceedings of the Fifth International Workshop on Software Specification and Design: 169-176.

Leite J.S.P., Freeman P.A. 1991. Requirements Validation through Viewpoint Resolution. IEEE Transactions on Software Engineering 17(12):1253-1269.

Levenshtein V.I. 1966. Binary Codes Capable of Correcting Deletions, Insertions and Reversals. Doklady Akademii Nauk SSSR 10(8): 707-710.

Ocampo A, Boggio D, Münch J, Palladino G. 2003. Toward a Reference Process for Developing Wireless Internet Services. IEEE Transactions on Software Engineering 29(12): 1122-1134.

Robinson W.N. 1989. Integrating Multiple Specifications Using Domain Goals. Proceedings of the Fifth International Workshop on Software Specification and Design: 219-226.

Turgeon J, Madhavhji H.N. 1996. A Systematic, View-Based Approach to Eliciting Process Models. EWSPT05: Proceedings of the European Workshop on Software Process Technology: 276-282.

Verlage M. 1998. An Approach for Capturing Large Software Development Processes by Integration of Views Modeled Independently. SEKE: Proceedings of the Tenth International Conference on Software Engineering and Knowledge Engineering: 227-235

**Figure 1. Empirically based approach for creating software development process models.**

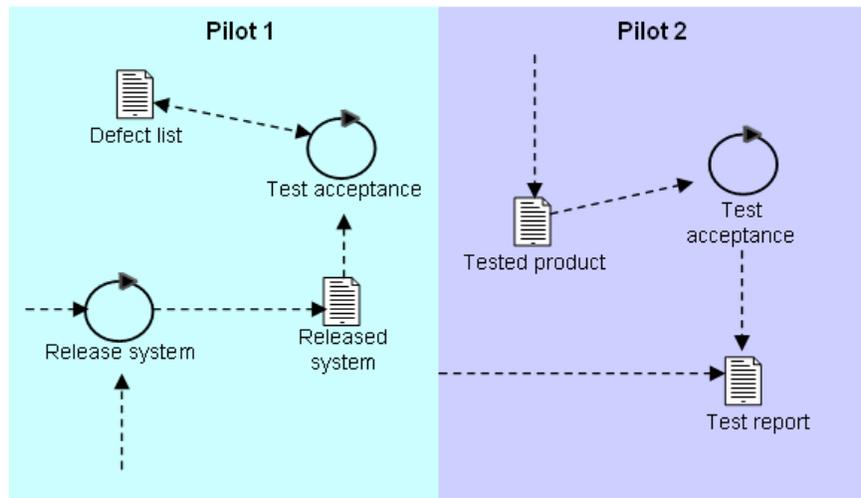

**Figure 2. Manual commonality analysis (excerpt).**

**Table 1. Characterization vectors**

| Customization factor | Characteristic | Project 1 | Project 2 |
|---|---|---|---|
| Domain characteristics | Application type | Information system | Computation intensive system |
| | Business area | Mobile online trading services | Mobile online entertainment services |
| Development characteristics | Project type | Client System adaptation | Client New development Server New development |
| | Transport protocol | GPRS/UMTS | GPRS/UMTS |
| | Implementation language | WML, J2ME | Client: J2ME Server: J2EE |
| Enterprise characteristics | Organizational context | Investnet-Italy | Motorola GSM- Italy VTT- Finland |
| | Role | Service provider, content provider, service developer | Technology provider, service developer |

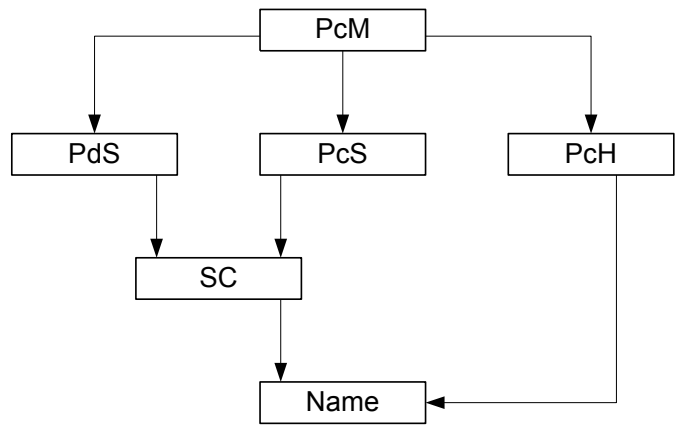

**Figure 3. Rules overview.**

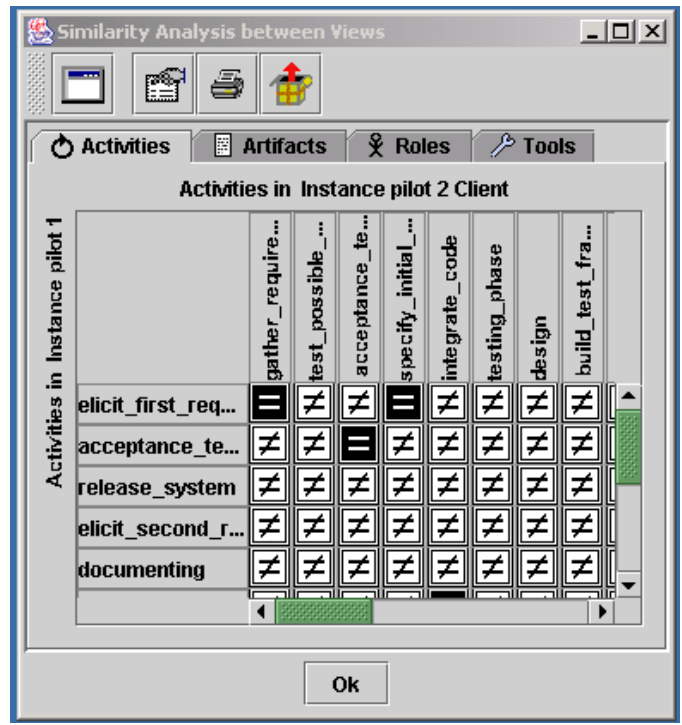

**Figure 4. Commonalities table (excerpt)**

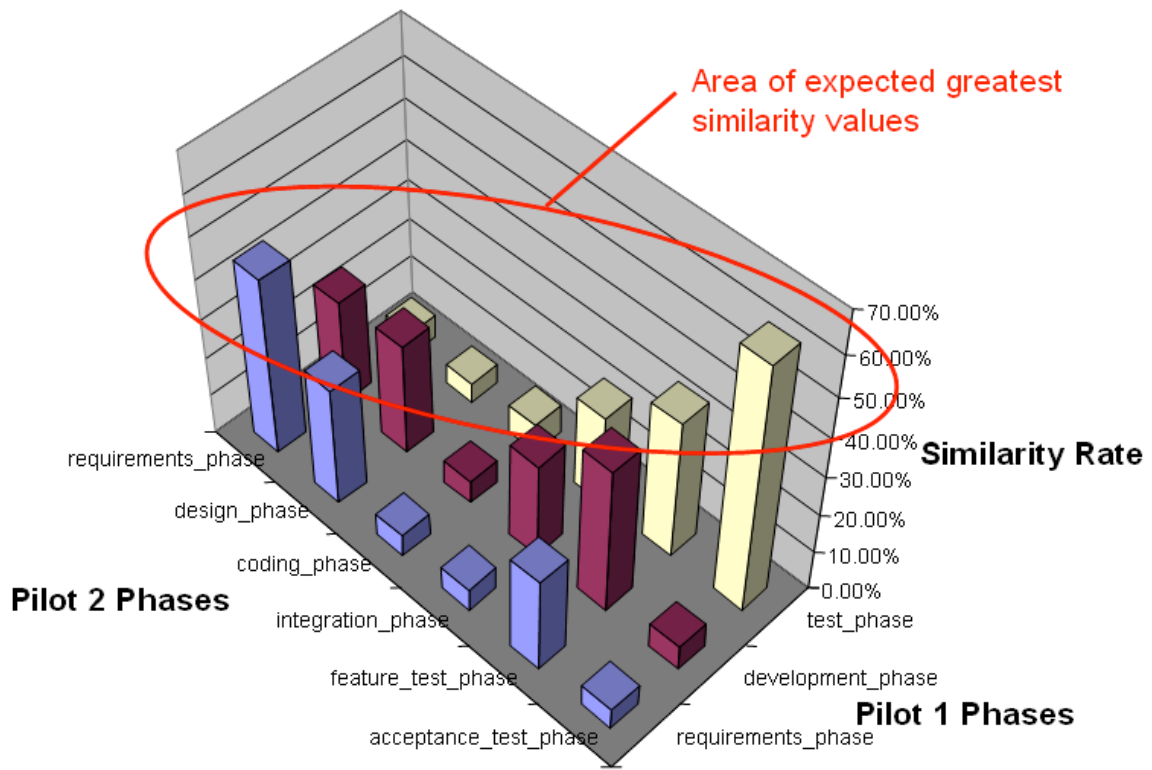

**Figure 5. Commonality values among phases.**

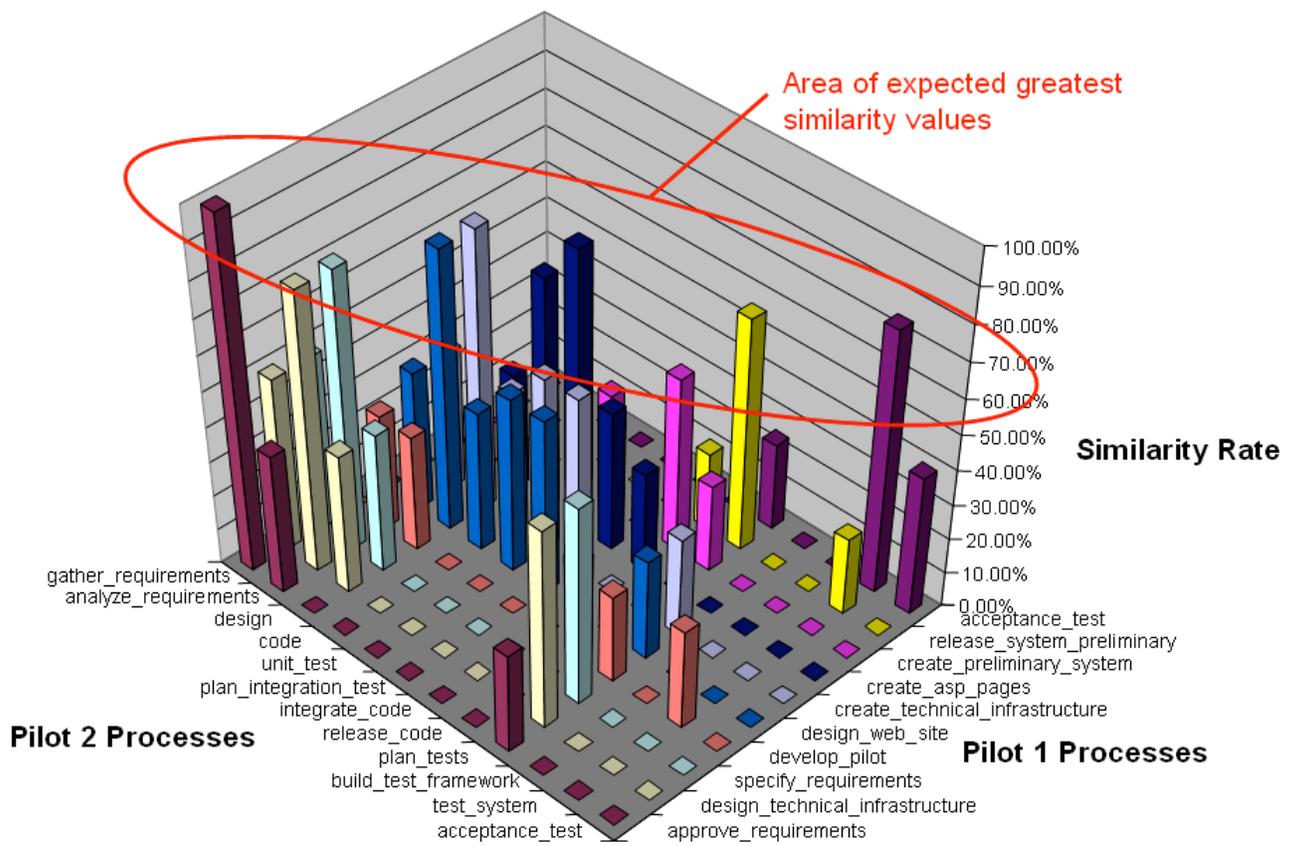

**Figure 6. Commonalities among technical processes.**

**Table 2. Entities and Icons**

| Entity | Icon |
|---|---|
| Alternative Box | 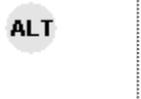 |
| Optional Box | 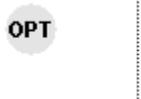 |

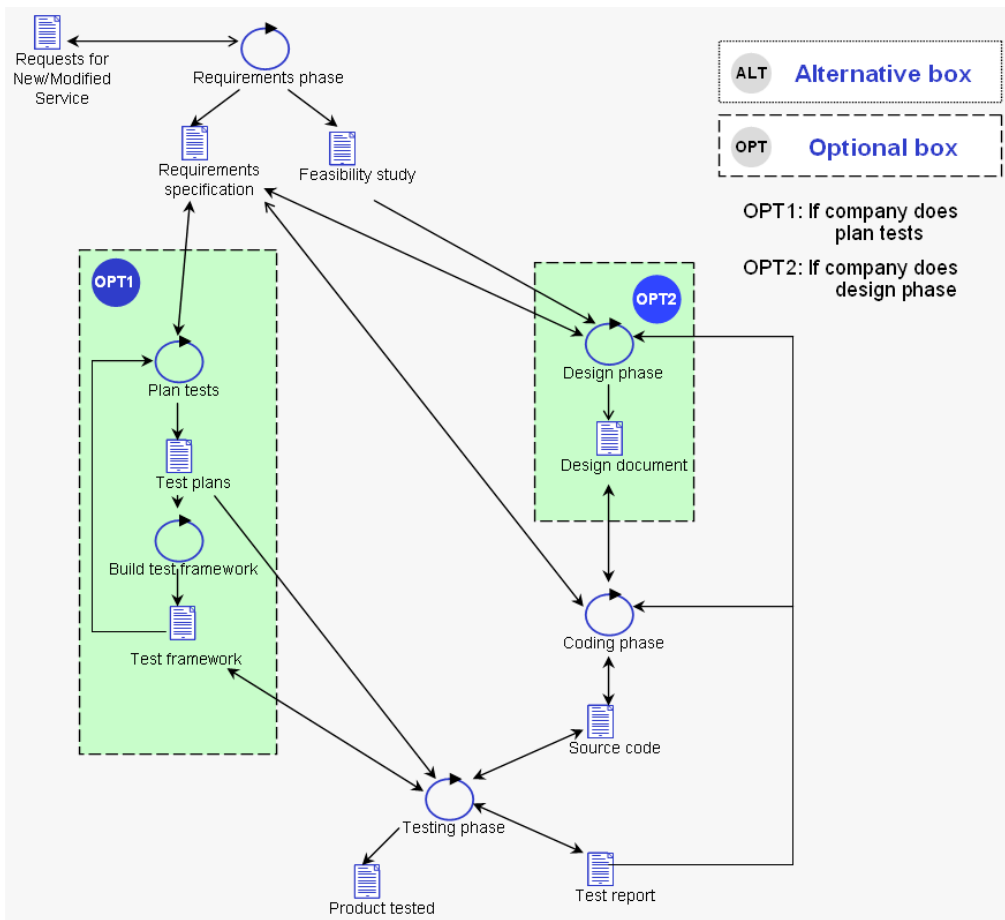

**Figure 7. Reference Process (excerpt).**

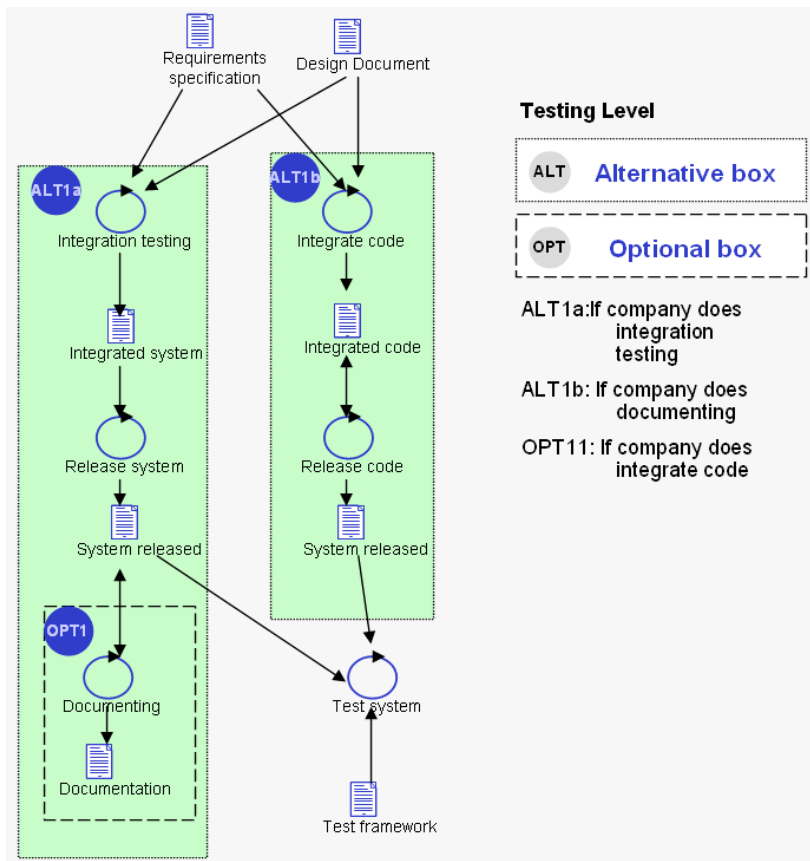

**Figure 8. Integrating/Releasing Process.**